\documentclass[twocolumn]{article}
\usepackage[margin=0.6in,columnsep=0.15in]{geometry}
\usepackage{graphicx} %
\usepackage[utf8]{inputenc}
\usepackage{multicol}
\usepackage{bm}
\usepackage{amssymb} 
\usepackage{amsmath} 
\usepackage{csquotes}
\usepackage{float}
\usepackage{stfloats}
\usepackage{gensymb}
\usepackage{chemformula}
\usepackage{orcidlink}
\usepackage{hyperref}
\usepackage{braket}
\usepackage{multirow}
\usepackage[normalem]{ulem}
\usepackage[LGR,T1]{fontenc}
\usepackage{textalpha}
\usepackage{layouts}
\usepackage[url=false, sorting=none]{biblatex}
\newcommand{\boldtheta}{{\boldsymbol\theta}}

\usepackage{authblk}

\begin{document}

\title{\Large An improved penalty-based excited-state variational Monte Carlo approach with deep-learning ansatzes}

\author[1,$^\circ$]{P. B. Szab\'o}
\author[1,$^\circ$]{Z. Sch\"{a}tzle}
\author[1]{M. T. Entwistle}
\author[1,2,3,4,*]{F. No\'{e}}
\affil[1]{FU Berlin, Department of Mathematics and Computer Science,
Arnimallee 6, 14195 Berlin, Germany}
\affil[2]{Microsoft Research AI4Science, Karl-Liebknecht Str. 32, 10178 Berlin, Germany}
\affil[3]{FU Berlin, Department of Physics, Arnimallee 14, 14195 Berlin, Germany}
\affil[4]{Rice University, Department of Chemistry, Houston, TX 77005, USA}

\date{}

\twocolumn[{%
  \maketitle
  \vspace{-3em}
  \begin{center}
  \begin{minipage}{0.85\linewidth}
    \small
    \paragraph{Abstract}
    We introduce several improvements to the penalty-based variational quantum Monte Carlo (VMC) algorithm for computing electronic excited states of Entwistle \textit{et al.} [M. T. Entwistle \textit{et al.}, Nat. Commun. \textbf{14}, 274 (2023)], and demonstrate that the accuracy of the updated method is competitive with other available excited-state VMC approaches.
A theoretical comparison of the computational aspects of these algorithms is presented, where several benefits of the penalty-based method are identified.
Our main contributions include an automatic mechanism for tuning the scale of the penalty terms, an updated form of the overlap penalty with proven convergence properties, and a new term that penalizes the spin of the wave function, enabling the selective computation of states with a given spin.
With these improvements, along with the use of the latest self-attention-based ansatz, the penalty-based method achieves a mean absolute error below 1 kcal/mol for the vertical excitation energies of a set of 26 atoms and molecules, without relying on variance matching schemes.
Considering excited states along the dissociation of the carbon dimer, the accuracy of the penalty-based method is on par with that of natural-excited-state (NES) VMC, while also providing results for additional sections of the potential energy surface, which were inaccessible with the NES method.
Additionally, the accuracy of the penalty-based method is improved for a conical intersection of ethylene, with the predicted angle of the intersection agreeing well with both NES-VMC and multi-reference configuration interaction.
  \end{minipage}
  \end{center}
  \vspace{1em}
}]

\makeatletter
\def\blfootnote{\xdef\@thefnmark{}\@footnotetext}
\makeatother

\blfootnote{$^*$~Email: \href{mailto:frank.noe@fu-berlin.de}{frank.noe@fu-berlin.de}}%
\blfootnote{$^\circ$~Equal contribution}%

\makeatletter
\if@twocolumn
\newcommand{\whencolumns}[2]{
#2
}
\else
\newcommand{\whencolumns}[2]{
#1
}
\fi
\makeatother

\section{Introduction}
    The central challenge towards an ab-initio description of chemical processes is solving the electronic Schrödinger equation for molecules and materials. 
    Its solutions provide, in principle, a full description of a system's electronic properties, facilitating simulation from first principles.
    While many quantum chemistry methods target only electronic ground states, access to low-lying excited states is necessary for accurately modeling phenomena such as the interaction of light and matter or catalysis \cite{matsika2018,lu2021,kancherla2019,stirbet2020,pena2003}.
    Light-matter interactions are of utmost importance for many processes in photochemistry, such as photoisomerization in the retinal chromophore \cite{schoenlein1991} or photodissociation in photosynthesis \cite{vasilev2001}.
    These photo-induced processes are notably at the core of several research frontiers, including the enhancement of light-harvesting materials for solar cells \cite{curutchet2017,kundu2017} and the development of photo-triggered drugs and medical screening devices \cite{fomina2012,bogomolny2009}.

    Despite the demand for accurate approximations of electronic excited states, their simulation remains challenging to this date \cite{mai2020,gonzalez2012,lischka2018}.
    While density functional theory is the workhorse of many quantum chemistry simulations, its applicability to excited states is limited and its extensions, such as time-dependent density functional theory (TDDFT), have their well-known limitations \cite{ullrich2011,liang2022,laurent2013}.
    Single-reference coupled cluster theory can be extended to target excited states, but it fails to correctly describe bond breaking and struggles with degeneracies, requiring a costly multi-reference treatment \cite{laurent2013}.
    A commonly used alternative for the computation of excited states is the complete active space self-consistent field (CASSCF) algorithm. 
    Although CASSCF has demonstrated considerable success, it scales exponentially with the size of the active space.
    Consequently, it relies heavily on a careful selection of active orbitals, often requiring chemical intuition and prior knowledge of the system under investigation \cite{keller2015}.
    Furthermore, the choice of compatible active spaces across molecular geometries poses additional complications for modeling excited-state potential energy surfaces.
    As a consequence, the modeling of excited-state dynamics often involves substantial human intervention, relying heavily on chemical intuition and a trial-and-error approach.

    Variational quantum Monte Carlo (VMC) provides a promising alternative to the established protocols. 
    In recent years the introduction of neural-network wave functions has significantly enhanced the accuracy of VMC \cite{hermann2020,pfau2020}. 
    The application of neural-network wave functions for the VMC simulation of molecular systems has proven highly successful in accurately describing states with multi-reference character and modeling strong correlation \cite{hermann2023}.
    Recently, these methods have been successfully extended to enable simulations of excited states \cite{entwistle2023} via a penalty-based formalism. %
    While results on some applications have already been promising, the methodology for simulation of excited states with neural-network based VMC is still under active development \cite{liu2023,pfau2024,wheeler2024, lu2023}.
    In this work, we introduce advancements in penalty-based VMC for excited states and conduct a comparative analysis of our approach with alternative methods, such as the recently proposed natural excited state quantum Monte Carlo (NES-VMC) \cite{pfau2024}, a sequential variant of the penalty-based method \cite{liu2023}, as well as the auxiliary wave function approach of Lu \textit{et al.} (AW) \cite{lu2023}.

    Our main contribution is the greatly improved accuracy of the penalty-based method for excited states, due to improvements in the optimization and the use of the more expressive self-attention-based Psiformer wave function architecture \cite{glehn2022}. 
    The methodological improvements include the alteration of the loss function of Entwistle \textit{et al.}  \cite{entwistle2023} to a form with proven convergence properties \cite{wheeler2024}, and automatically tuning the scale of the penalty term to reduce noise in the gradients, while retaining the global minimum of the loss function and preventing the collapse of the states. 
    We utilize the KFAC optimizer \cite{martens2015} and tune its hyperparameters to account for the number of model parameters increasing with the number of excited states. 
    Furthermore, we introduce a method to target states with a specific spin by combining the selection of the magnetic quantum number of the spin-assigned wave function with a spin penalty that favors low-spin states.
    This approach enhances the efficiency of computing excited states by segmenting them into spin sectors and enables precise targeting of desired spin states.
    The method is combined with the use of pseudo potentials \cite{annaberdiyev2018,wang2019}, which we employ for heavy atoms (second row and beyond) throughout this work.
    We present results for our improved method on a variety of atomic and molecular systems ranging from 4 to 42 electrons. 
    The first set of experiments focuses on single-point calculations, demonstrating high accuracy on the ten lowest-lying states of first- and third-row atoms, as well as on the five lowest-lying states of a variety of organic and inorganic molecules.
    In order to validate the accuracy of our wave functions, we compute the oscillator strengths of ground- to excited-state transitions in the molecules.
    The second part of the experiments targets excited-state potential energy surfaces, where we model the intricate electronic structure of the carbon dimer and the conical intersection of ethylene.
    For the carbon dimer, we recover a large fraction of the potential energy surface for a total of nine states, composed of four singlet, four triplet and a quintet state.
    For the ethylene isomerization process, we improve over previous calculations with penalty-based excited-state VMC and predict the pyramidalization angle of the conical intersection in good agreement with the NES-VMC method. 
    
\section{Method}
The aim of wave function-based electronic structure methods is to approximate the solution to the time-independent electronic Schrödinger equation:
\begin{equation}
    \hat H \Psi^i = E^i \Psi^i \, ,
\end{equation}
where $\hat H$ is the electronic Hamiltonian with eigenstates $\Psi^i$ and corresponding energies $E^i$, with the states conventionally ordered according to increasing energy.
For molecular systems in the Born--Oppenheimer approximation and using atomic units, $\hat H$ takes the form
\begin{equation}\label{eq:hamil}
    \hat H = \frac{1}{2} \sum_i \nabla_i^2 - \sum_{iI} \frac{Z_I}{\mathbf{R}_{iI}} + \sum_{i<j} \frac{1}{\mathbf{r}_{ij}} \, ,
\end{equation}
with $Z_I$ being the nuclear charges, while $\mathbf{R}_{iI}$ and $\mathbf{r}_{ij}$ denote electron-nucleus and electron-electron distances, respectively.
As the electronic Hamiltonian is Hermitian, all $E^i$ are real and $\Psi^i$ are orthogonal to each other.
Furthermore, we assume all $\Psi^i$ to be real-valued which can always be ensured for molecular wave functions.
Ground-state electronic structure approaches are concerned by computing only the lowest energy $E^0$, and corresponding ground-state wave function $\Psi^0$ of a given system.
The objective of excited state methods extends to computing the lowest $n$ energies along with their corresponding states for the system under consideration.

\subsection{Variational optimization}
\subsubsection{Variational Monte Carlo}
Variational quantum Monte Carlo (VMC) belongs to the family of variational methods for approximating the eigenstates of a quantum many-body system. 
These methods are based on the variational principle of quantum mechanics, which states that the functional given by the expectation value of the Hamiltonian
\begin{equation}
    \mathcal{L}[\Psi] = \braket{\hat{H}}_\Psi = \frac{\braket{\Psi|\hat{H}|\Psi}}{\braket{\Psi|\Psi}}\,,
\end{equation}
is minimized for the (unnormalized) ground-state wave function $\Psi^0$. 
This Rayleigh quotient can be understood as a loss function for optimization of a parameterized ansatz $\Psi_\boldtheta$:
\begin{equation}
    \boldtheta^* = \underset{\boldtheta}{\text{argmin}} ~ \mathcal{L}[\Psi_\boldtheta] \,,
\end{equation}
where $\boldtheta$ are the model parameters and the optimal parameters $\boldtheta^*$ are to be found.
A challenging aspect of this optimization is the computation of integrals over the high-dimensional domain of the wave function. 
VMC solves this problem by numerically approximating integrals, such as the computation of the expectation value of observable $\hat{O}$, through Monte Carlo integration:
\begin{equation}\label{eq:montecarlo}
    \braket{\hat{O}}_{\Psi_\boldtheta} \approx \frac{1}{n} \sum^n_{{\bf r} \sim |\Psi_{\boldtheta}|^2}\frac{\hat{O}\Psi_\boldtheta({\bf r})}{\Psi_\boldtheta({\bf r})}\,,
\end{equation}
outlining an algorithm that poses very few restrictions on the form of ansatz parametrization. 
Minimizing Monte Carlo estimates of the objective function amounts to an alternating scheme of sampling electron positions ${\bf r} \in \mathbb R^{3N}$ from the probability density associated with the square modulus of the wave function $\Psi_\boldtheta$ and updating the model parameters $\boldtheta$ with a variant of stochastic gradient descent.
Note that for the electronic Hamiltonian it is convenient to work with the spin-assigned real-valued wave function $\Psi_\boldtheta({\bf r}): \mathbb R^{3N} \rightarrow \mathbb R$.
For a detailed description of the VMC method with application to neural-network wave functions see the work of Sch\"{a}tzle \textit{et al.}\cite{schatzle2023}

\subsubsection{Optimizing excited states with the overlap penalty}\label{sec:overlap-penalty}
The variational method can be extended to excited states by accounting for the orthogonality of the eigenstates. 
The relevant partition of the spectrum of the Hamiltonian can be constructed by finding the orthogonal subspace with minimal energy.
A conceptually simple way of approximating the $n$ lowest eigenstates $\{\Psi^0,...,\Psi^n\}$ of a system is to impose their orthogonality by extending the objective function with an overlap penalty:
\begin{equation}
    \mathcal{L}[\Psi_{\boldtheta}^0, ..., \Psi_{\boldtheta}^n] = \sum_i \braket{\hat{H}}_{\Psi_{\boldtheta}^i} +  \sum_{i<j} \alpha_{ij} |\braket{\Psi_{\boldtheta'}^i | \Psi_{\boldtheta}^j}|^2 \,,
\end{equation}
resulting in the coupled optimization of multiple ansatzes.
The global minimum of this objective function is attained for the lowest-lying states, if $\alpha_{ij}$ is chosen to be larger than the energy gap between the $i$th and $j$th state \cite{wheeler2024}. 
The overlap between each pair of (unnormalized) states can again be computed by Monte Carlo integration \cite{entwistle2023}. 
To enforce pure eigenstates and increase the training stability, we apply the orthogonality constraint only with respect to lower-lying states, resulting in a total of $\tfrac{n(n-1)}{2}$ relevant overlap terms, for $n$ states.
To achieve this, the gradients of the overlap term are computed only with respect to $\Psi^j_{\boldtheta}$ by detaching $\Psi^i_{\boldtheta'}$ from the computational graph.
Note that in contrast to the above construction, a loss function with symmetric overlap penalty terms yielding gradients to both the lower and higher lying states would be minimized by all linear combinations of the lowest $n$ states.
While Entwistle \textit{et al.}\cite{entwistle2023} used a variant of the loss function that diverges upon the collapse of two eigenstates, increasing training stability at the cost of a small bias in the excitation energies, our improved method is unbiased, and stable using the simpler penalty term introduced by Pathak \textit{et al.}\cite{pathak2021} 
Recently Wheeler \textit{et al.} \cite{wheeler2024} have introduced an ensemble method, where the presented loss function can be derived in a more general framework.

\subsubsection{Targeting spin states with spin penalty}
In many applications, selection rules prohibit excitations that would involve changing the spin of the electronic state.
Consequently, states within a fixed spin sector are often of interest. 
While states of the targeted spin sector can be selected from a simulation of all low-lying eigenstates through evaluation of the spin of the acquired wave functions, this procedure potentially involves the computation of many ultimately irrelevant states. 
A common approach to address this issue is to offset states based on their spin, pushing them out of the targeted region of the spectrum.
We employ a similar technique in the context of VMC by augmenting our loss function with a spin penalty:
\whencolumns{
    \begin{equation}
        \mathcal{L}[\Psi_{\boldtheta}^0, ..., \Psi_{\boldtheta}^n] = \sum_i \braket{\hat{H}}_{\Psi_{\boldtheta}^i} + \beta\sum_i \braket{\hat{S}^2}_{\Psi_{\boldtheta}^i} + \sum_{i<j} \alpha_{ij} |\braket{\Psi_{\boldtheta'}^i | \Psi_{\boldtheta}^j}|^2 \,,\
        \label{eq:excited_loss}
    \end{equation}
}{
    \begin{equation}
    \begin{split}
        \mathcal{L}[\Psi_{\boldtheta}^0, ..., \Psi_{\boldtheta}^n] = & \sum_i \braket{\hat{H}}_{\Psi_{\boldtheta}^i} + \beta\sum_i \braket{\hat{S}^2}_{\Psi_{\boldtheta}^i} + \\
        & \sum_{i<j} \alpha_{ij} |\braket{\Psi_{\boldtheta'}^i | \Psi_{\boldtheta}^j}|^2 \,,\
        \label{eq:excited_loss}
    \end{split}
    \end{equation}
}
where $\braket{\hat{S}^2}_{\Psi^i}$ is the expectation value of the squared magnitude of the spin operator and $\beta$ weighs the penalty term.
The expectation value of the spin operator is evaluated through Monte Carlo sampling, as described in Section (\ref{sec:observables}).
As the molecular Hamiltonian commutes with the spin operator, they share a common set of eigenstates.
This makes pure spin states a valid target of the variational optimization and evaluation of the spin expectation value with Monte Carlo integration efficient. 
For sufficiently large $\beta$, this objective function favors solutions with low total spin, that is singlet (doublet) states for systems with an even (odd) number of electrons. 
In order to obtain higher spin states, i.e. triplet (quartet) states, we fix the $m_S$ component of our ansatz accordingly. 
We again include the spin penalty, now yielding the solution minimizing the total spin magnitude within the subset of constrained wave functions.
Although the restriction to the subspace of higher spin states by fixing the difference between spin-up and spin-down electrons has been previously discussed in the context of penalty-based excited state optimization \cite{liu2023}, the integration of the spin penalty enables us to leverage this concept to target states with specific spins.
For a detailed explanation of the treatment of spin in VMC wave functions, see Supporting Information Section S3. 
Note that while minimizing the total spin of the wave function allows for reformulating the gradients to contain only first derivatives \cite{entwistle2023}, this is no longer possible when targeting a specific value of the spin expectation through the minimization of the difference from that targeted value.
Furthermore, we highlight that the aforementioned penalty approach can be extended to restrict VMC ansatzes based on other observable quantities. 
For example, it's possible to penalize or favor states of a certain spacial symmetry or target other properties of the eigenstates if their operators commute with the Hamiltonian.

\subsection{Comparison with other VMC methods for excited states.}\label{sec:method_comparison}
While the field of VMC for molecules in first quantization using neural-network wave functions has mostly settled on the optimal strategy for ground-state optimization, the methodology of excited-state optimization is still subject to ongoing development. 
In this section, we compare the penalty-based method with other available strategies and point out some of their respective advantages and disadvantages.

The most recent alternative to penalty-based VMC for excited states is the natural-excited-state VMC (NES-VMC) approach of Pfau \textit{et al.} \cite{pfau2024}.
In NES-VMC, the problem of finding the lowest $n$ eigenstates of the physical Hamiltonian is transformed to finding the ground state of the Hamiltonian of an extended system with $n$ times as many electrons.
Unlike penalty-based approaches, this elegant formulation allows trial wave functions to be optimized to match a linear combination of the lowest excited states without imposing additional constraints.
The main drawback of NES-VMC lies in the necessity to model and sample an extended system of electrons.
The larger effective system size increases computational costs in two ways.
Firstly, it requires the evaluation of much larger determinants, which constitutes the step with the steepest theoretical scaling in the whole deep-learning VMC algorithm.
Secondly, wave functions of the extended system form a higher dimensional Hilbert-space than those of the physical system, leading to a quicker onset of the ``curse of dimensionality''.
On top of the increased computational costs, NES-VMC requires an additional diagonalization of the local energy matrix to recover the energies and wave functions of the individual states from the solution of the extended system.
This also means that one does not have access to the energies and other properties of the states throughout optimization, making the convergence of individual states difficult to measure without incurring further computational costs for frequent diagonalizations.
A small additional drawback of the coupled nature of the states in NES-VMC is the requirement for heuristics to prevent numerical instabilities when states become nearly linearly dependent.
Lastly, unlike in penalty-based VMC, there is no straightforward way to impose restrictions on the spin of the ansatzes.
This limitation may require the simulation of other, irrelevant, lower-lying states, simply to access the higher-lying states of interest.

Next, the sequential variant of the penalty-based method for excited states recently proposed by Liu \textit{et al.} \cite{liu2023} is considered.
In this approach, a simple ground-state computation is carried out to convergence before running a second calculation for the first excited state.
The second calculation incorporates an overlap penalty term with respect to the fixed ground state obtained from the first calculation.
The process is then repeated for higher-lying excited states, with a growing number of penalty terms applied to all previously obtained wave functions.
While this strategy might help stabilize the early steps of training and potentially enable access to higher-lying states when optimizing numerous ansatzes in parallel is prohibitively expensive, its sequential nature makes it difficult to use in practice.
Most importantly, one only receives feedback on the quality of the highest state after completing the computations for all preceding states.
This could result in a significant amount of repeated computation compared to regular penalty-based VMC, where the quality of all states can be assessed from the earliest stages, allowing for early adjustments.
Additionally, the necessity to fully converge all computations for the lower-lying states before starting the next calculation means there is no flexibility to stop the training early if relative energies have converged, or to continue training if the desired accuracy has not yet been achieved, without breaking the sequential paradigm.

Another closely related alternative is the auxiliary wave function method of Lu \textit{et al.} \cite{lu2023}.
This method avoids the dependence on a free parameter to scale the overlap penalty by combining the orthogonalization of Choo \textit{et al.} \cite{choo2018} with the real-space overlap computations used by Entwistle \textit{et al.}\cite{entwistle2023}.
In this method, excited states are found by running sequential optimizations where previously converged lower-lying states are projected out from the currently optimized ansatz.
To achieve this, overlaps between the current wave function and all lower-lying wave functions are computed, allowing the energy contributions from the $n-1$ lower-lying states to be subtracted, to obtain the energy expectation value of the $n$th excited state.
While this method eliminates the hyperparameter used for weighting the overlap penalty, it does so at the cost of retaining only an implicit representation of the excited-state wave functions, while still requiring the computation of the same overlap terms as in the penalty-based method.
Additionally, the computation of observables necessitates evaluating all pairwise overlaps with respect to each lower-lying state, leading to significantly higher computational costs and making the targeting of specific spin states infeasible.
Furthermore, stochastic errors in the difficult-to-estimate overlap terms affect the evaluation of observables more directly than in penalty-based VMC.
Lastly, Lu \textit{et al.} demonstrated that their results can be reproduced with penalty-based methods, incurring slightly lower computational costs, if the overlap penalty is weighted correctly.
This task is facilitated by the automatic penalty scaling introduced in this work.

Finally, we consider the method of targeting excited states via variance minimization \cite{umrigar1988,umrigar2005}.
This approach minimizes the variance of the local energies of each state, leveraging the fact that this variance should be zero for all eigenstates of the Hamiltonian.
The method relies on sufficiently accurate initial guesses to ensure convergence to the desired states.
Cuzzocrea \textit{et al.} \cite{cuzzocrea2020} demonstrated that for increasingly complex Slater--Jastrow type ansatzes, the variance minimization scheme becomes prone to escape the local minima of the target state, and often converges to other states with lower variance.
A further complication of variance minimization with neural-network ansatzes is the appearance of mixed third derivatives of the wave function in the gradient expression of the loss function.
Due to the large number of ansatz parameters, these quantities significantly increase the computational cost of the gradient calculation, which already forms one of the bottlenecks of the algorithm.
Variance minimization has therefore not been often employed in the recent neural-network-based VMC approaches to excited states.

\subsection{Neural-network wave function ansatz}
Neural-network-based wave function ansatzes have been very successful in describing intricate many-body correlation in quantum systems \cite{hermann2023}. 
The state-of-the-art architectures for molecules in first quantization are implemented via the concept of linear combinations of generalized Slater determinants:
\begin{equation}\label{eq:dl_ansatz}
    \psi_{\boldtheta}(\mathbf r_1, ..., \mathbf r_N) =
  \textstyle{\sum}_p
  \det[\mathbf A^p(\mathbf r)] \, ,
\end{equation}
employing the determinant as an antisymmetrizer over permutation equivariant many-body orbitals:
\begin{equation}\label{eq:mb_orbital}
    A^p_{ik}= \phi^p_k(\mathbf r_i, \{\mathbf r^\uparrow\},\{\mathbf r^\downarrow\}) \times \varphi^p_k(\mathbf r_i) \, .
\end{equation}
The expressivity of the wave functions arises from the parametrization of $\phi^p_k$ as neural networks, while the envelope functions $\varphi^p_k$ implement the correct asymptotics. 
The major difference between the existing architectures is how the latent-space representation of the electrons, ultimately projected to obtain the many-body orbitals $\phi^p_k$, is constructed from the electron and nuclear positions. 
The experiments throughout this work are performed with the Psiformer architecture \cite{glehn2022}. 
The Psiformer electron embeddings $\mathbf{h}$ are instantiated based on the (scaled) electron nuclei distances and their respective spin.
Electronic correlation is then built up incrementally through subsequent self-attention interactions, resembling the encoder part of a transformer:
\whencolumns{
    \begin{align}
        \mathbf{h}_i^{l+1} &= \mathbf{f}_i^{l+1} + (\mathbf{W}^{l+1}\mathbf{f}^{l+1} + \mathbf{b}^{l+1}), \\
        \mathbf{f}_i^{l+1} &= \mathbf{h}_i^{l} + \mathbf{W}_o^{l}\bigoplus_h[\textsc{SelfAttn}_i(\mathbf{h}_1^{l},...,\mathbf{h}_N^{l}; \mathbf{W}_{q}^{lh},\mathbf{W}_{k}^{lh},\mathbf{W}_{v}^{lh})]\,,
    \end{align}
}{
    \begin{align}
        \mathbf{h}_i^{l+1} =& \, \mathbf{f}_i^{l+1} + (\mathbf{W}^{l+1}\mathbf{f}^{l+1} + \mathbf{b}^{l+1}), \\
    \begin{split}
        \mathbf{f}_i^{l+1} =& \, \mathbf{h}_i^{l} + \\
        & \mathbf{W}_o^{l}\bigoplus_h[\textsc{SelfAttn}_i(\mathbf{h}_1^{l},...,\mathbf{h}_N^{l}; \mathbf{W}_{q}^{lh},\mathbf{W}_{k}^{lh},\mathbf{W}_{v}^{lh})]\,,
    \end{split}
    \end{align}
}
where $\textsc{SelfAttn}$ is the standard multi-headed self-attention block \cite{vaswani2017}, $\bigoplus_h$ denotes the concatenation over attention heads $h$, $i$ indexes electrons and $l$ indexes the layer. The subscripts $q,v,k$ stand for queries, values and keys, respectively and $\mathbf{W}$ and $\mathbf{b}$ are the weights and biases of the Psiformer layer.
An electron-wise linear projection on vectors of dimension $N\cdot N_{det}$ is applied to transform the output of the last layer to the orbitals $\phi^p_k$.
Additionally, the electronic cusps are modeled with a multiplicative Jastrow factor:
\whencolumns{
    \begin{equation}
        J(\mathbf r_1, ..., \mathbf r_N) = \sum_{i<j; \sigma_i=\sigma_j} - \frac{1}{4}\frac{\alpha_p^2}{\alpha_p + |\mathbf{r}_i-\mathbf{r}_j|} \; +
        \sum_{i,j; \sigma_i\ne\sigma_j} -\frac{1}{2}\frac{\alpha_a^2}{\alpha_a + |\mathbf{r}_i-\mathbf{r}_j|} \, .
    \end{equation}
}{
    \begin{equation}
    \begin{split}
        J(\mathbf r_1, ..., \mathbf r_N) = & \sum_{i<j; \sigma_i=\sigma_j} - \frac{1}{4}\frac{\alpha_p^2}{\alpha_p + |\mathbf{r}_i-\mathbf{r}_j|} \; + \\
        & \sum_{i,j; \sigma_i\ne\sigma_j} -\frac{1}{2}\frac{\alpha_a^2}{\alpha_a + |\mathbf{r}_i-\mathbf{r}_j|} \, .
    \end{split}
    \end{equation}
}

The Psiformer is implemented in the \textsc{DeepQMC} program package \cite{hermann2023a}.
For more information on the Psiformer we refer to the original publication of von Glehn \textit{et al.} \cite{glehn2022}, while further details about the implementation of neural-network wave functions in \textsc{DeepQMC} can be found in Schätzle \textit{et al} \cite{schatzle2023}.
Lastly, we emphasize that the penalty-based method for computing excited states can be applied in combination with any other valid ansatz architecture as well.

\subsection{Pretraining and CASSCF baseline}
It is well known \cite{pfau2020,schatzle2023} that molecular ground-state VMC calculations employing neural-network ansatzes greatly benefit from a short supervised pretraining phase preceding the variational optimization.
In this stage, a mean squared error loss function between the many-body orbitals of the ansatz and the single-particle orbitals of a reference HF or CASSCF solution is minimized:
\whencolumns{
\begin{equation}
    \mathcal{L}^\text{pre}= \frac{1}{n} \sum_i \left\langle 
        \sum_{pjk} \left(
            \varphi^{\text{ref},p}_{k}({\bf r}_j) - \phi^p_k({\bf r}_j, \{{\bf r}^\uparrow\}, \{{\bf r}^\downarrow\}) \times \varphi_k({\bf r}_j)
        \right)^2
    \right\rangle_{\Psi^i_\boldtheta} \, ,
\end{equation}
}
{
\begin{equation}
\begin{split}
    \mathcal{L}^\text{pre} & = \frac{1}{n} \sum_i \Big\langle 
    \sum_{pjk} \big(
        \varphi^{\text{ref},p}_{k}({\bf r}_j) \\
    &- \phi^p_k({\bf r}_j, \{{\bf r}^\uparrow\}, \{{\bf r}^\downarrow\}) \times \varphi_k({\bf r}_j)
    \big)^2
    \Big\rangle_{\Psi^i_\boldtheta} \, ,
\end{split}
\end{equation}
}
where $\varphi^{\text{ref},p}_{k}$ are the orbitals of the reference solution with $p$ enumerating determinants and $k$ the respective orbitals.
This pretraining serves as an informed initialization scheme for the parameters of the neural network, and can help in avoiding convergence to local minima, along with reducing the number of expensive variational optimization steps necessary to achieve a certain threshold of accuracy.

In the context of excited-state calculations, one can use the lowest $n$ roots of a multi-state CASSCF calculation as a pretraining target.
Care must be taken during the definition of the active space \cite{entwistle2023}, such that the resulting Slater-determinants of the CASSCF solution have the right spin configurations, and contain the necessary orbitals to describe all excitations of interest.
The latter is of special importance when a great number of excited states of the smallest systems are considered, such as for the lithium atom in Section \ref{sec:atoms}.
In the case of the lithium atom, the lowest states can all be described as excitations of the single valence electron to higher and higher orbitals.
To qualitatively describe the lowest $n$ states, one must include at least $n$ orbitals in the active space, which in turn might necessitate the use of single particle basis sets larger than the ones usually employed in deep-learning VMC pretraining targets.
In the present work, the relatively large aug-cc-pVTZ \cite{dunning1989a,kendall1992a,wilson1999a} basis set is used for most systems to ensure a quantitatively correct initialization of the highest excited states considered.
The only exception is benzene, where the cc-pVDZ basis set \cite{wilson1999a} is used instead, as the CASSCF computation with the aug-cc-pVTZ basis set is deemed too expensive, and where there is no need for diffuse basis functions.
When targeting states with a specific spin, this has to be reflected in the calculation of the CASSCF baseline by restricting to the selected spin sector.
The number of pretraining iterations is set to one thousand for all systems except for benzene, where 100 000 pretraining iterations were used, to make the variational training as efficient as possible \cite{glehn2022}.
All CASSCF calculations have been performed with the \textsc{PySCF} program package \cite{sun2020}.
Minimal active spaces are chosen for all systems in the present study, selected by searching for the smallest active spaces that still produce sensible excitation energies with respect to the reference values.
While these calculations give a qualitatively valid picture of the excitation energies, they are far from the quantitative accuracy of the deep-learning VMC simulations.
For example, the mean absolute error of the excitation energies predicted by the baseline CASSCF calculations for the atoms and molecules considered in Section \ref{sec:single_point} is 350 meV (8 kcal/mol).

\subsection{Evaluating observables}\label{sec:observables}
In quantum mechanics, the wave function gives a complete description of the state of the system. 
Having access to the electronic wave function of molecules grants theoretical access to all their observable electronic properties. 
To extract these properties from our wave function models, Monte Carlo integration is employed.
For single state properties equation \eqref{eq:montecarlo} can be applied directly.
For off-diagonal properties between unnormalized states, such as the overlap or the transition dipole moment, we follow Entwistle \textit{et al.} \cite{entwistle2023} and evaluate the geometric mean of the Monte Carlo estimates with respect to either of the two wave functions.
In the following, we sketch out the evaluation of the expectation values of the total spin magnitude and the oscillator strength operators.

\subsubsection{Spin magnitude}
The total spin magnitude of the spin-assigned wave function manifests in the symmetries of its spatial part (see Supporting Information Section S3). It can therefore be obtained from the spin-assigned wave function, by evaluating its symmetry properties under exchanges of opposite-spin particles \cite{huang1998}.
We follow the procedure employed in NES-VMC \cite{pfau2024}, and evaluate the spin as:
\whencolumns{
\begin{equation}
    \braket{\hat{S}^2}_{\Psi_\boldtheta} = - \frac{N_\uparrow-N_\downarrow}{4}(N_\uparrow-N_\downarrow+2)+N_\downarrow-\Big\langle\sum_{1\le i\le N_\uparrow}\sum_{N_\uparrow<j\le N}\frac{\Psi_\boldtheta(...,\mathbf{r}_j,...,\mathbf{r}_i,...)}{\Psi_\boldtheta(...,\mathbf{r}_i,...,\mathbf{r}_j,...)}\Big\rangle_{\Psi_\boldtheta} \, ,
\end{equation}
}{
\begin{equation}
\begin{split}
    \braket{\hat{S}^2}_{\Psi_\boldtheta} = &- \frac{N_\uparrow-N_\downarrow}{4}(N_\uparrow-N_\downarrow+2)+N_\downarrow\\
    &-\Big\langle\sum_{1\le i\le N_\uparrow}\sum_{N_\uparrow<j\le N}\frac{\Psi_\boldtheta(...,\mathbf{r}_j,...,\mathbf{r}_i,...)}{\Psi_\boldtheta(...,\mathbf{r}_i,...,\mathbf{r}_j,...)}\Big\rangle_{\Psi_\boldtheta} \, ,
\end{split}
\end{equation}
}
where $N^\uparrow$ ($N^\downarrow$) denotes the number of spin-up (spin-down) electrons, respectively.
While the evaluation of the spin scales with $N_\uparrow \cdot N_\downarrow \sim (\tfrac{N}{2})^2$, it does not involve the computation of the local energy and is therefore of negligible cost for the systems of the present study.

\subsubsection{Transition dipole moment and oscillator strength}
The oscillator strength is a useful quantity to approximate the rate of transitions between electronic states as a result of interaction with light, under the constant electric field and dipole moment assumptions.
Previous studies on neural-network-based VMC for excited states have used the accuracy of the predicted oscillator strength as a proxy to assess the quality of the underlying wave function models \cite{entwistle2023, pfau2024}.
This is motivated by the fact that the oscillator strength is highly dependent on the quality of the wave function approximation and thus provides a useful measure for the accuracy of the wave function on top of the energy. The oscillator strength $f_{ij}$ is a dimensionless quantity obtained from the energy gap $\Delta E_{ij}$ and the absolute value of the transition dipole moment defined as the expectation of the dipole operator $\hat{\boldsymbol\mu} = \sum_iq\hat{\mathbf{r}_i}$ between states $i$, and $j$:
\begin{equation}
    f_{ij}=\frac{2}{3}\Delta E_{ij}|\braket{\Psi^i_\boldtheta|\hat{\boldsymbol{\mu}}|\Psi^j_\boldtheta}|^2\,.
\end{equation}
In practice, the off-diagonal expectation value of the dipole operator is evaluated using reweighted samples from the wave functions, and the excitation energy $\Delta E_{ij}$ is obtained by independently evaluating the energy expectation values of the two states.

\subsection{Implementation details}
This section describes the most important technical details of the penalty-based method as implemented in the \textsc{DeepQMC} program package \cite{schatzle2023}.
In \textsc{DeepQMC}, the conventional deep-learning-based ansatzes used for ground-state calculations are conveniently extended to model a range of electronic states via the \textsc{JAX} framework's \cite{jax2018github} vectorizing map function transformation.
Using this transformation, the $\mathbb{R}^{3N} \to \mathbb{R}$ ansatz functions parametrized by a single set of parameters $\boldtheta$ become functions $\mathbb{R}^{n\times3N} \to \mathbb{R}^n$ parametrized by $n$ sets of parameters $[\boldtheta^0, ..., \boldtheta^n]$.
In combination with the excited-state loss function of (\ref{eq:excited_loss}), these extended ansatzes can then be used in the existing deep-learning VMC framework of \textsc{DeepQMC}, with a few minor modifications detailed below.
To obtain the $n$ sets of electron configuration samples in parallel, the same vectorizing map transformation is used on the singe-state Markov-chain Monte Carlo sampling routines.
The hyperparameters of the KFAC optimizer \cite{martens2015} used during the variational training are adjusted as follows.
Since an ansatz that describes $n$ electronic states is parametrized by $n$ times as many parameters as a single-state ansatz, the squared norm of a parameter update will on average be $n$ times larger as well ($\sum_n \sum_i (\Delta\boldtheta^n_i)^2$ compared to $\sum_i (\Delta \boldtheta_i)^2$).
Accordingly, the KFAC hyperparameter controlling the maximum squared norm of parameter updates is scaled by the number of computed states, $n$.
The effect of this change is most pronounced on the largest considered systems, such as benzene, where the larger absolute values of total energies can yield gradients with larger magnitudes, while on most of the smaller systems investigated here the effect is not noticable.

Lastly, the treatment of the free parameters $\alpha_{ij}$ deserves some attention.
As demonstrated by Pathak \textit{et al.} \cite{pathak2021}, and later refined by Wheeler \textit{et al.} \cite{wheeler2024} these hyperparameters can be chosen freely without affecting the global minimum of the loss function, as long as
\begin{equation}
    \alpha_{ij} > |E^j - E^i| \, .
    \label{eq:alpha_constraint}
\end{equation}
Furthermore, we've found that while all $\alpha_{ij}$ values satisfying (\ref{eq:alpha_constraint}) avoid the collapse of the optimized states, values closer to the $| E^j - E^i |$ limit can in some cases reduce the noisiness of the training.
Unfortunately, choosing the optimal value for $\alpha_{ij}$ while satisfying the above constraint can require system-dependent manual tuning of these parameters.
To alleviate this issue, the automatic scaling of the $\alpha_{ij}$ parameters based on running estimates of $E^i$ and $E^j$ is introduced.
In particular, the scaling of the overlap penalty between states $i$ and $j$ is computed in each training step as
\whencolumns{
\begin{equation}
\alpha_{ij} = \tilde \alpha \cdot \text{max}\left(| \text{ewm}(\bar E_\text{loc}^j) - \text{ewm}(\bar E_\text{loc}^i) |, \, \text{ewm}(\sqrt{\text{Var}(E_\text{loc}^i)}), 10^{-3} \; \text{E}_h\right) \, ,
\end{equation}
}
{
\begin{equation}
\begin{split}
    \alpha_{ij} = \tilde \alpha \cdot \text{max}\Big(&| \text{ewm}(\bar E_\text{loc}^j) - \text{ewm}(\bar E_\text{loc}^i) |,\\
    &\text{ewm}(\sqrt{\text{Var}(E_\text{loc}^i)}), 10^{-3} \; \text{E}_h\Big) \, ,
\end{split}
\end{equation}
}

where $\tilde \alpha > 1$ is the new free parameter shared between all pairs of states, $\text{ewm}(\cdot)$ denotes the exponentially weighted mean over the training iterations, $E_\text{loc}$ is the batch of local energies in the current step, overbar denotes the mean, and $\text{Var}(\cdot)$ the variance.
The first argument of the maximum function ensures that the constraint (\ref{eq:alpha_constraint}) is fulfilled while automatically scaling $\alpha_{ij}$ in a system-specific way, while the second argument prevents the collapse of the states in the earliest stages of the training where $\text{ewm}(\bar E_\text{loc}^i)$ is not yet a good estimate of $E^i$.
As a result of this parametrization, the optimal value for the new parameter $\tilde \alpha$ is significantly less system-dependent than that of $\alpha_{ij}$.
For the bulk of the systems considered here, $\tilde \alpha$ is set to four to ensure a comfortable margin of safety in satisfying (\ref{eq:alpha_constraint}), while for a handful of systems it is decreased to two and one to ensure optimal convergence in all cases.
For concrete values of $\tilde \alpha$ employed for each system, see the Supporting Information.
It is recognized that reducing the dependence on the value of the $\tilde \alpha$ parameter warrants further research.
On the other hand, considering the already limited range in which the parameter is varied, along with the numerous benefits of the method compared to other excited-state VMC approaches (detailed in Section \ref{sec:method_comparison}), the penalty-based method can already be considered relevant, accurate, and easy-to-apply in practice.

\subsection{Scaling}
\begin{figure}[bp!]
    \centering
    \includegraphics{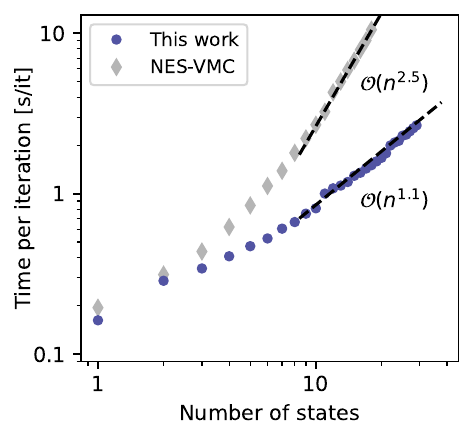}   
    \caption{\textbf{Scaling of the computational cost with the number of states.} The wall-clock time of a single training iteration is shown for the neon atom, with an electron configuration batch size of 64 on a single A100 GPU. Dashed lines represent least-squares fits to the penalty-based and NES-VMC results with $n \ge 10$.}
    \label{fig:scaling}
\end{figure}
The two dominant factors determining the cost of any excited-state VMC computation are the size of the considered physical system, and the number of computed electronic states.
A clear advantage of the penalty-based method over extended-system approaches such as NES-VMC is that these two axes of scaling are almost entirely decoupled from each other.
In the NES-VMC method, increasing the system size and considering more electronic states both increase the number of simulated fermions, bringing with it the usual difficulties including reduced sampling efficiency, the need to compute determinants of larger matrices, and the worsening of the ``curse of dimensionality''.
In contrast, considering new electronic states in penalty-based VMC only adds fermions that are simulated almost independently from the ones already present, interacting with them only through the overlap penalty term of (\ref{eq:excited_loss}).
\begin{figure}[bp!]
    \centering
    \includegraphics{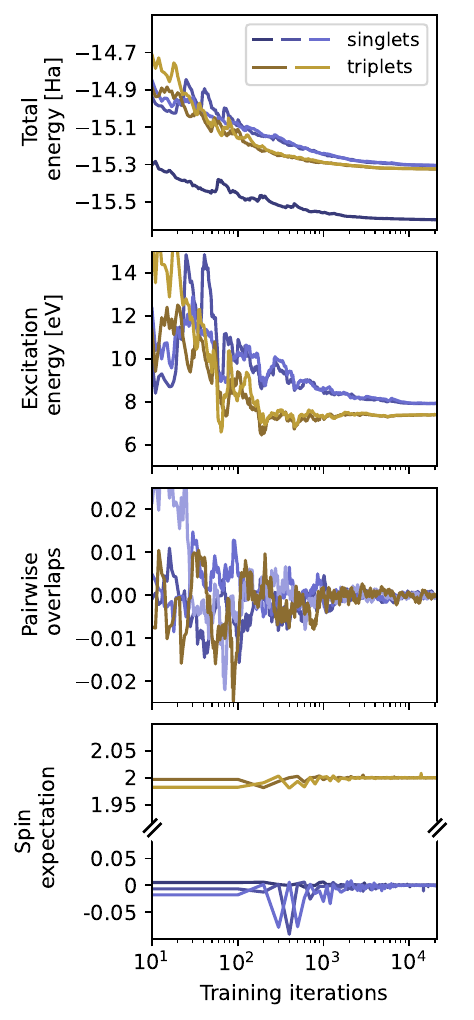}
    \caption{{\bf Convergence of relevant quantities throughout the optimization.}
    Two penalty-based excited-state VMC computations are carried out to characterize the lowest three singlet and two triplet electronic states of the HCl molecule.
    The plotted quantities from top to bottom: total energies, energies of the excitations from the ground state, pairwise overlaps $\langle \Psi_i \vert \Psi_j \rangle$, and the expectation values of the spin magnitude operator $\langle \hat S^2 \rangle_{\Psi_i}$.}
    \label{fig:convergence}
\end{figure}
In fact, the computation of this overlap term is the only part of the algorithm that scales quadratically with the number of electronic states, while the time complexity and memory requirement of all other steps scale linearly.
Fortunately, the cost of the overlap computation is much smaller than that of the local energies, when considering 1--30 electronic states, and therefore the penalty-based method exhibits very favorable scaling in this regime, as demonstrated in Figure \ref{fig:scaling}.
The points on this plot are obtained by performing a small number of training iterations for the neon atom using an electron configuration batch size of 64, while varying the number of computed states. 
To reduce noise each experiment has been averaged over three repetitions.
With ten states or more, one finds an empirical scaling of roughly $\mathcal{O}(n^{1.1})$, indicating a practically linear scaling of the penalty-based method with the number of electronic states.
Considering the NES-VMC results on the same system, one finds a much steeper approximate scaling of $\mathcal{O}(n^{2.5})$ which is, as expected, similar to the scaling of ground-state deep-learning VMC methods with the number of electrons \cite{schatzle2023}.
This steep scaling results in the fact that while a single state iteration takes roughly the same time with the NES-VMC code \cite{githubFermiNet}, an iteration with ten states already takes four times as long.
Lastly, we note that when during inference with the penalty-bassed method, one is interested in the properties of only a subset of the states (e.g the highest-lying states only), one is free to decouple the optimized ansatzes, and evaluate expectation values only for the selected states.
This can help avoiding unnecessary computation compared to the NES-VMC method, where the ansatzes of the different states are only meaningful as a single unit, and cannot be easily decoupled from each other.
\section{Results}
\begin{table*}[tp]
\centering
\caption{Accuracy comparison between various excited-state VMC methods on vertical excitation energies of the systems in the present study.
The last three columns contain the mean absolute errors, maximum errors, and standard deviation of errors in the excitation energies,  respectively, for the given method in units of meV.
Results of the sequential variant of the penalty-based method are taken from  Liu \textit{et al.} \cite{liu2023}, penalty-based numbers with the PauliNet ansatz are reproduced from Entwistle \textit{et al.} \cite{entwistle2023}, while NES-VMC results are taken from Pfau \textit{et al.}\cite{pfau2024}
Partitioning of the systems into subsets is necessary as the various works considered slightly different sets of atoms and molecules.\vspace{0.3cm}}
\begin{tabular}{lll|ccc}
 &  &  & MAE & MAX & STD \\
Test set & Subset & Method & [meV] & [meV] & [meV] \\
\hline\hline
\multirow[c]{7}{*}{Atoms} & \multirow[c]{2}{*}{First row atoms} & NES-VMC & 17 & 263 & 37 \\
 &  & This work & 37 & 166 & 43 \\
 \cline{2-6}
  & \multirow[c]{2}{*}{Li--O, 10 states} & Sequential PB & 39 & 107 & 39 \\
 &  & This work & 31 & 140 & 37 \\
  \cline{2-6}
 & \multirow[c]{2}{*}{Li--O, 5 states} & PauliNet & 85 & 518 & 120 \\
 &  & This work & 21 & 85 & 32 \\
 \cline{2-6}
 & \multirow[c]{2}{*}{Ge--Se} & Sequential PB & 28 & 50 & 25 \\
 &  & This work & 21 & 64 & 25 \\
 \cline{2-6}
 & All atoms & This work & 33 & 166 & 40 \\
 \hline
\multirow[c]{9}{*}{Molecules} & \multirow[c]{2}{*}{without BeH} & NES-VMC & 34 & 284 & 67 \\
 &  & This work & 45 & 176 & 60 \\
 \cline{2-6}
 & \multirow[c]{2}{*}{without BeH, C\textsubscript{6}H\textsubscript{6}} & NES-VMC & 33 & 284 & 69 \\
 &  & This work & 32 & 90 & 37 \\
 \cline{2-6}
  & \multirow[c]{2}{*}{BH, H\textsubscript{2}O, CO} & PauliNet & 134 & 258 & 103 \\
 &  & This work & 27 & 58 & 26 \\
\cline{2-6}
 & BeH, CO, H\textsubscript{2}O, & Sequential PB & 73 & 153 & 86 \\
 & H\textsubscript{2}S, H\textsubscript{2}CSi & This work & 24 & 58 & 23 \\
 \cline{2-6}
 & All molecules & This work & 45 & 176 & 60 \\
\end{tabular}
\label{tab:single_point_stats}
\end{table*}
\subsection{Convergence of the penalty-based optimization}

Before turning to the various benchmark test sets, it is instructive to consider the convergence of the relevant quantities throughout the penalty-based excited-state VMC computation, in order to gain an intuitive understanding of the method.
In Figure \ref{fig:convergence}, the evolution of the total energy, the excitation energy, the pairwise overlaps and the spin expectation value are depicted throughout a VMC optimization for the lowest five states of the HCl molecule. 
The lowest excited states of HCl are twofold degenerate triplet states followed by a set of doubly degenerate singlet states, leading to several interesting characteristics of the molecular spectrum.
The spin penalty term is utilized, and two separate simulations are performed for the singlet (3 states) and triplet (2 states) spin sector. 
The ansatzes for the triplet states are assigned two unpaired spin-up electrons, ensuring that these wave functions will have $m_z=1$.
As usual, the variational optimization is preceded by a short supervised pretraining. 
It can be seen that all three terms entering the loss function converge smoothly to their optimal values. 
Due to the pretraining, the states start off approximately orthogonal and within the correct spin sector, which is maintained throughout the optimization.
The excitation energies converge rapidly and stabilize after approximately ten thousand training iterations.
Similar training trajectories are obtained for the other experiments throughout this paper.
For a comprehensive evaluation of the optimized wave function properties, the training is followed by an evaluation stage, during which the observables of interest are sampled extensively with fixed wave function parameters.
Post-processing of the wave function, such as the diagonalization step in the NES-VMC and AW methods, is not required.

\subsection{Single point calculations}\label{sec:single_point}
\begin{figure*}[t]
    \centering
    \includegraphics{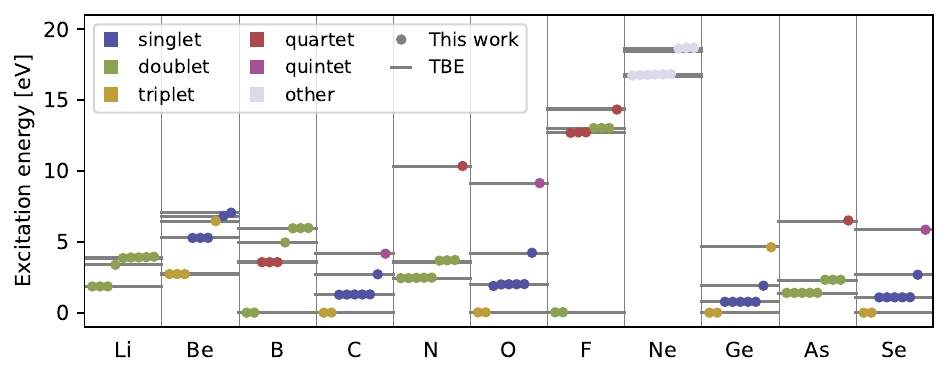}
    \caption{{\bf Excitation energies of first- and third-row atoms.}
    The lowest nine electronic excitation energies are computed for all first-row, and a range of third-row (Ge--Se) atoms.
    Experimental reference values from the NIST Atomic Spectra Database are depicted with gray horizontal lines \cite{NIST_ASD}, while dots display the results of the present work, with colors denoting the spin of the given excited state.}
    \label{fig:atoms}
\end{figure*}

\subsubsection{Atoms}\label{sec:atoms}
In this section, we demonstrate the capability of the penalty-based excited-state method to model an extended number of excited states simultaneously by computing the lowest nine excitation energies of a range of first- and third-row atoms (lithium to neon and germanium to selenium).
We employ highly accurate, experimentally determined atomic spectral lines as reference, after removing the effect of spin-orbit coupling by weighted averaging of the finely split levels.
For the first-row atoms, all electrons are included in the computations, while for the third-row atoms, the electrons occupying the three lowest shells are replaced with the ccECP pseuodopotential \cite{bennett2017}.

The computed excitation energies are plotted in Figure \ref{fig:atoms}, alongside the reference experimental data.
It is clear that these excitation energies provide a quantitatively correct description of all states in question.
To further asses the accuracy of our method we examine the mean absolute errors (MAEs) of the excitation energies for the atoms obtained with different methods in Table \ref{tab:single_point_stats}.
We find that on the subset of first row atoms NES-VMC exhibits an advantage with 17 meV (0.39 kcal/mol), compared to the 37 meV (0.86 kcal/mol) of penalty-based VMC.
The impressive accuracy of NES-VMC was achieved \cite{pfau2024} by performing twice as many training iterations as is done in the present work.
It is plausible that the accuracy of the penalty-based method would continue to improve similarly with additional training steps.
Considering the distribution of errors across the different excitation energies, it becomes apparent that the NES-VMC method tends to accumulate most of its error in the highest computed excited state.
In contrast, the penalty-based method offers an advantageous, relatively uniform description of all states.
With a maximum error of 166 meV (3.8 kcal/mol), the penalty-based method appears to suffer less from the occasional outliers present in the NES-VMC results, which exhibit a maximum error of 263 meV (6.1 kcal/mol).
Inspecting the individual states more closely, the errors of the sharpest outliers in the NES-VMC results (the highest computed states of boron and fluor) are improved by more than a factor of two and six, respectively, in the present work.
On the contrary, compared to NES-VMC, the penalty-based method appears to encounter slightly more difficulty with highly degenerate states, such as the first quintuple degenerate excited state of nitrogen.
This is presumably due to the accumulating noise from the numerous terms in the loss function, derived from the overlaps between states that are very close in energy.
Consequently, this noise in the gradients of the loss hinders the convergence of the degenerate states, leading to overestimated excitation energies.
This issue is somewhat mitigated by scaling the weight of these loss terms by the energy difference between the given states, as described in Section \ref{sec:overlap-penalty}, to achieve a MAE of 72 meV (1.7 kcal/mol) for these five excitation energies.
While we generally found the variance matching scheme introduced alongside the first version of penalty-based VMC for excited-states \cite{entwistle2023} to not be necessary, it can help improving the accuracy for highly degenerate states.
If one were to apply this variance matching, the MAE for the five degenerate excitation energies of nitrogen would be reduced to just 14 meV (0.34 kcal/mol).
Fortunately, electronic states with this level of degeneracy are exceedingly rare in molecular systems outside of single atoms, therefore we don't expect this to pose a serious limitation to the method in practice.
Lastly, we note that as opposed to the first published version of penalty-based excited-state VMC with neural-network ansatzes \cite{entwistle2023}, here all states of the lithium atom are found correctly.
Comparing now with the sequential variant of the penalty-based method for excited states \cite{liu2023} on the first row atoms from lithium up to oxygen, the present work achieves a favorable MAE of 31 meV (0.72 kcal/mol) compared to 39 meV (0.90 kcal/mol).
The additional error likely arises from the use of pseudopotentials, particularly in combination with the lightest atoms, where the energies of the two core electrons are not sufficiently separated from the valence energies for the pseudopotential approximation to be fully valid.
Turning to the third-row atoms, where both works employ pseudopotentials, one finds slightly lower MAEs of 21 meV (0.48 kcal/mol) and 28 meV (0.64 kcal/mol) for the present and the sequential penalty-based results, respectively.
Considering the results on the atomic systems overall, the updated penalty-based VMC method appears capable of describing as many as ten, often highly degenerate electronic states simultaneously, at a high level of accuracy.

\subsubsection{Molecules}
\begin{figure*}[t]
    \centering
    \includegraphics{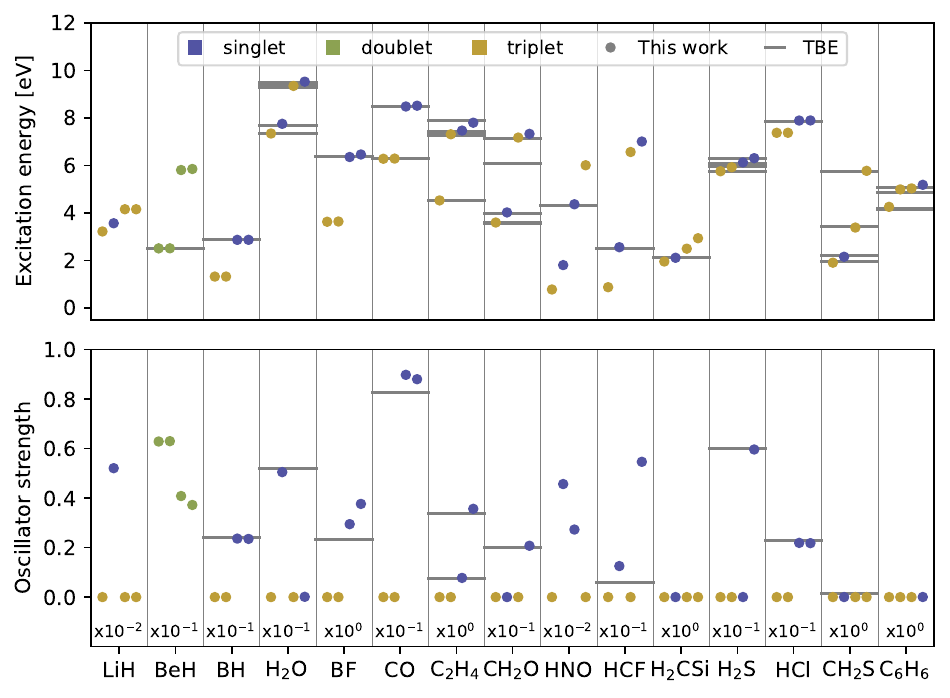}
    \caption{{\bf Excitation energies and oscillator strengths of main-group molecules.} Electronic excitation energies and oscillator strengths are computed for the lowest four transitions in a set of 15 molecules.
    Theoretical best estimates are depicted with gray horizontal lines where available \cite{veril2021}, while the predictions of penalty-based VMC are displayed with dots, with colors denoting the spin of the corresponding excited state.
    Excitation energies are plotted on the top pane, while oscillator strengths are shown on the bottom pane.}
    \label{fig:molecules}
\end{figure*}
Next, the five lowest-lying excited states of fifteen molecules are considered, with system sizes ranging from 4 to 42 electrons.
The geometries of LiH and BeH are taken from Entwistle \textit{et al.} \cite{entwistle2023}, benzene is taken from the work of Loos \cite{loos2020} \textit{et al.}, while the remaining twelve geometries are taken from Chrayteh \textit{et al.} \cite{chrayteh2021}.
The latter two publications also provide a number of accurate excitation energies in the complete basis set limit, and oscillator strengths in a triple-zeta quality basis, both obtained with high-order coupled cluster methods, which serve as references here.
The statistical measures of accuracy for molecules displayed in Table \ref{tab:single_point_stats} are computed only for the states where reference data is available (see the horizontal lines of Figure \ref{fig:molecules}).
Electrons from the first two shells of second-row atoms are replaced with the ccECP potential \cite{annaberdiyev2018}.
It should be noted that, especially in the case of oscillator strengths, reference values obtained with different basis sets or at different orders of the coupled cluster expansion may exhibit wider disparities than the differences between the investigated QMC methods, highlighting the difficulty of reliably estimating these quantities.
\newpage
On the top pane of Figure \ref{fig:molecules}, the first four vertical excitation energies of the fifteen molecules in question are plotted.
Similar to the findings for atoms, we observe that both the NES and penalty-based methods yield excitation energies in very good agreement with the theoretical best estimates.
As demonstrated in Table \ref{tab:single_point_stats}, when compared to the atomic test set, the MAEs of the two methods are somewhat closer for the molecules, measuring at 34 meV (0.78 kcal/mol) for NES-VMC and 45 meV (1.0 kcal/mol) for penalty-based VMC.
This closing of the accuracy gap can largely be attributed to the absence of highly degenerate states in molecules, which were the primary cause of noisy and outlier states in the atomic systems.
In terms of errors made on individual molecules and states, the penalty-based excited-state method shows improvement over the potentially misconverged highest tioformaldehyde state of NES-VMC.
Additionally, it performs slightly better on the third excitation energy of nitroxyl.
However, it does make larger errors on benzene, as well as on the highest states of formaldehyde and ethylene.
It is worth noting that despite the two methods exhibiting very similar MAEs, their error distributions are quite different.
The predictions of the NES-VMC method exhibit slightly lower errors than those of the penalty-based method for the majority of systems, but at the same time contain more severe outliers for a handful of states.
In contrast, the penalty-based approach tends to produce relatively uniform errors, which could conceivably be further reduced by conducting more training iterations.
This trend is also evident in the slightly larger standard deviation of errors for NES-VMC.
Interestingly, both VMC-based methods fail to capture the $^3A_1$ state of formaldehyde (both with FermiNet and Psiformer), potentially indicating a more fundamental issue with the description of this state, which warrants further research.
The fact that the failure to describe this state is reproducible across different ansatz architectures, loss functions, and pretraining schemes points to a more general optimization problem associated with this electronic state.
It is conceivable that the minimum of the loss function corresponding to this state is especially narrow or shallow, making it difficult to locate with stochastic gradient based optimization methods.
For convergence curves, reference and baseline CASSCF excitation energies obtained for this state see the Supporting Information.

Turning now to the oscillator strengths plotted on the bottom pane of Figure \ref{fig:molecules}, one finds a generally good agreement between the penalty-based excited-state VMC method and the reference results, with a few notable exceptions.
Specifically, the intensity of the highest two transitions of carbon monoxide and boron monofluoride are overestimated by penalty-based VMC compared to coupled cluster.
For certain excitations of the LiH, BeH, HNO, and HCF molecules, the lack of accurate reference data makes it difficult to ascertain the performance of the penalty-based excited-state VMC method.
Nonetheless, the generally accurate oscillator strength estimates obtained from the penalty-based method serve as compelling evidence that it not only delivers accurate energies but also well-converged wave functions, which can be of great use in computing a wide range of observable quantities of excited states.
Furthermore, it is clear that contrary to previous hypothesis \cite{pfau2024}, the penalty-based method is competitive with the NES-VMC approach when employing the same, sufficiently expressive ansatz in both cases.

\subsection{Excited-state potential energy surfaces}
\begin{figure*}[t]
    \centering
    \includegraphics{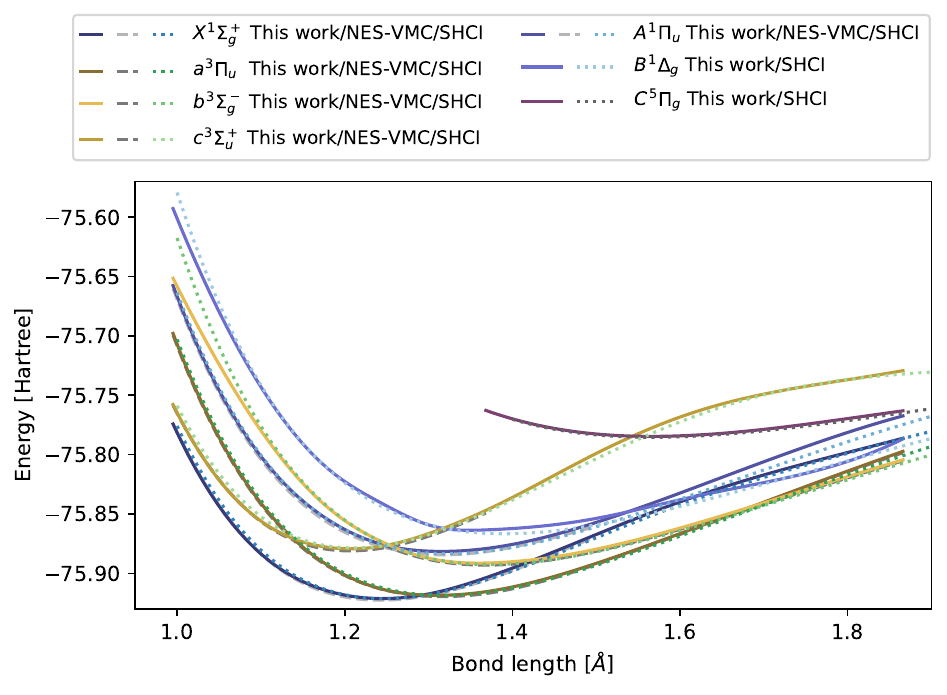}
    \caption{{\bf Excited potential energy surfaces of the carbon dimer.}
    Potential energy surfaces for nine of the low-lying excited states of the carbon dimer are calculated.
    Solid lines represent penalty-based excited-state VMC results, dashed curves were obtained using NES-VMC \cite{pfau2020}, and dotted lines depict semistochastic heat-bath configuration interaction results in the cc-pV5Z basis \cite{holmes2017}.}
    \label{fig:carbon-dimer}
\end{figure*}
While performing single-point computations targeting the excited states of molecules is a valuable endeavor in its own right, characterizing portions of molecular excited-state PESs holds even greater significance.
Access to these PESs can enable simulations of the evolution of molecular systems in their excited states \cite{cuzzocrea2022}, paving the way for describing crucial photo-chemical processes such as the light-harvesting step in photovoltaic devices \cite{cerullo2002,dahlberg2017} or the photoisomerization of the retinal chromophore that initiates vision \cite{g.rao2022}.
Furthermore, information about the PES is often essential for bridging theoretical results with experimental estimates, in tasks such as determining adiabatic excitation energies or the zero-point vibrational energy correction.
Unfortunately, the theoretical characterization of excited-state PESs often presents even greater challenges than a single-point calculation, such as the problem of consistently defining the active space in active-space-based methods, or the treatment of strong correlation near conical intersections in approximate time-dependent density functional theories \cite{malis2020,entwistle2023}.
In this section, we will demonstrate how the penalty-based method can be easily applied to compute excited-state molecular PESs for systems with challenging electronic structures.

\subsubsection{Excited states of the carbon dimer}
\begin{figure}[t]
    \centering
    \includegraphics{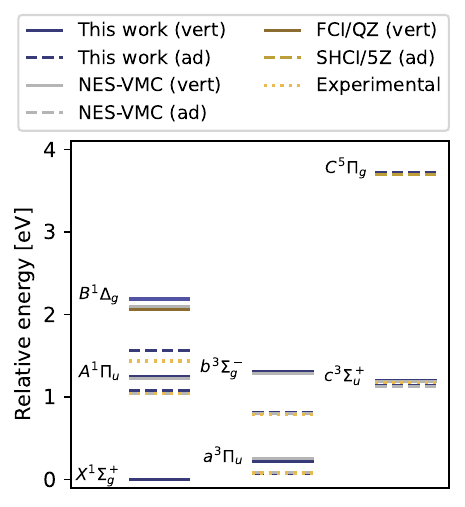}
    \caption{{\bf Vertical and adiabatic excitation energies of the carbon dimer.}
    Vertical (vert) excitation energies are plotted with solid, adiabatic (ad) ones with dashed, and experimental ones with dotted lines.
    The zero-point vibrational energy was not considered during the computation of the adiabatic excitation energies.
    The penalty-based VMC predictions are compared with results obtained from NES-VMC \cite{pfau2024}, full-configuration interaction using the cc-pVQZ basis \cite{loos2020}, semistochastic heat-bath configuration interaction with the cc-pV5Z basis \cite{holmes2017}, and experiments \cite{tanabashi2007}.}
    \label{fig:carbon-dimer-relative-energy}
\end{figure}
Despite its small size, the electronic structure of the carbon dimer remains the subject of intensive interest in both experimental \cite{martin1992} and theoretical studies \cite{sharma2015,holmes2017}.
It exhibits strong multi-reference character and numerous nearly-degenerate low-lying excited states with singlet, triplet and quintet spins, some of which can be characterized as double excitations.
Accurately and consistently describing its numerous electronic states across a range of bond lengths presents a challenge tackled only by the most sophisticated electronic structure methods.

The excited-state PESs of the carbon dimer, computed using the penalty-based excited-state method, are shown in Figure \ref{fig:carbon-dimer} with solid lines.
The capability of penalty-based VMC to perform separate computations for states with different spins has been leveraged to efficiently characterize the lowest four singlet, four triplet and one quintet states.
It is important to note that two of the singlet and two of the triplet states are degenerate across the entire range of bond lengths, resulting in only three-three lines being plotted for these spin sectors.
The results reported with NES-VMC \cite{pfau2024} are shown in Figure \ref{fig:carbon-dimer} with dashed lines.
In regions where NES-VMC results are available, they are in excellent agreement with the curves obtained from penalty-based excited-state VMC.
Note that the penalty-based method converges in multiple regions where its counterpart cannot provide a sufficiently accurate description, such as the compressed geometries of the $b^3\Sigma^-_g$ state, the stretched geometries of the $c^3\Sigma^+_g$ and all singlet states, as well as the entire $B^1\Delta_g$ curve.
Additionally, due to the computational efficiency afforded by the use of the spin penalty, one is able to describe a section of the lowest-lying quintet $C^5\Pi_g$ state, including its minimum, by performing a single-state penalty-based calculation.
Considering that both the $b^3\Sigma^-_g$ and $B^1\Delta_g$ states are characterized as double excitations \cite{pfau2024}, a class of excited states that many other methods struggle with, the ability of penalty-based VMC to accurately describe large sections of these PESs is particularly noteworthy.

Comparing the predictions of the penalty-based excited-state method with results obtained using the highly accurate stochastic heat-bath configuration interaction (SHCI) approach \cite{holmes2017} shown with dotted lines in Figure \ref{fig:carbon-dimer}, one finds excellent agreement for most geometries of all states.
Notable exceptions are the high-lying compressed regions of the $b^3\Sigma^-_g$ state, and the stretched regions of the $A^1\Pi_u$ and $B^1\Delta_g$ curves.
For the compressed geometries, the VMC-based method underestimates the excitation energies, while for the stretched geometries, it delivers slightly higher estimates.
In the stretched geometries of the singlet states, we observe higher-than-usual variance in the expectation of the energy during training, indicating potential problems with the fitting of these states, whereas we identified no such signals for the compressed triplet states.
Overall, the accuracy of the penalty-based method appears to be on par with that of NES-VMC and comparable with the SHCI reference.
Additionally, its consistency and ability to target specific spin states enable it to deliver results on larger section of the carbon dimer PESs than its VMC-based counterpart.

The vertical and adiabatic excitation energies from the ground state to the considered excited states of the carbon dimer are plotted in Figure \ref{fig:carbon-dimer-relative-energy}.
The relative energies obtained from penalty-based VMC are in good agreement with both NES-VMC and, where applicable, reference experimental \cite{tanabashi2007}, full configuration interaction \cite{loos2020}, and SHCI results \cite{holmes2017}.
The discrepancies between the penalty-based method and the appropriate references are well below 43 meV (1.0 kcal/mol) for all states except for $B^1\Delta_g$, where both the vertical and adiabatic excitation energies are overestimated by penalty-based VMC by about 100 meV (2.3 kcal/mol).

\subsubsection{Conical intersection in Ethylene}

\begin{figure*}[t]
    \centering
    \includegraphics{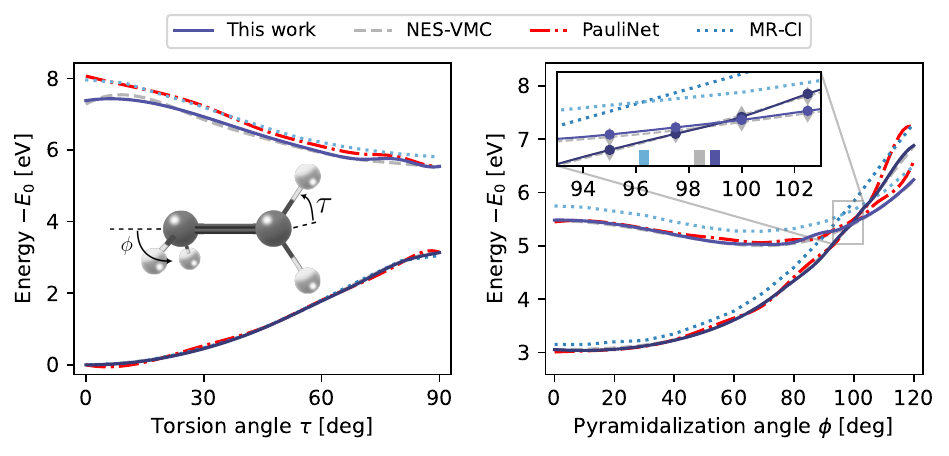}   
    \caption{\textbf{Conical intersection of ethylene.}
    The energies of the lowest two singlet states of ethylene are plotted as a function of the torsion and pyramidalization angles, relative to the energy of the ground state at the equilibrium geometry.
    Results obtained with the updated penalty-based method are plotted alongside those of the original version with the PauliNet ansatz \cite{entwistle2023}, NES-VMC \cite{pfau2024}, and multi-reference configuration interaction with single and double excitations \cite{barbatti2004}.
    The inset plot on the right pane shows the region around the conical intersection, with the vertical lines marking the approximate location of the conical intersection.}
    \label{fig:ethylene}
\end{figure*}

Conical intersections play an important role in the study of excited-state dynamics associated with processes such as photoisomerization and photodissociation, providing a pathway for non-radiative relaxation from electronic photoexcitations. 
From a computational perspective, conical intersections pose a significant challenge, as the multi-reference character of the electronic states increases when the potential energy surfaces converge. 
However, to achieve a good description of the dynamics, an accurate and well-balanced model of the excited-state potential energy surface is crucial.
Here we study the conical intersection of ethylene, which serves as a small-scale model system for photoswitches.
Upon photoexcitation to the lowest singlet excited state, ethylene undergoes torsion (angle $\tau$) along the C-C bond, followed by pyramidalization (angle $\phi$) of one of the CH$_2$ groups, as depicted in the inset of Figure \ref{fig:ethylene}.
This leads to a conical intersection, where a radiationless transition to the ground-state potential energy surface occurs \cite{barbatti2004}. 
The process involves intricate changes in the electronic structure that pose well-known problems for single-reference methods such as TDDFT \cite{schmerwitz2022}. 
While our focus lies on the study of the singlet states, the simulations are further complicated by the presence of multiple lower-lying triplet states. 

To address this, we employ the spin penalty method to restrict our calculations to the singlet sector, ensuring that no more than one excited state need to be computed at any point along the trajectory.  
The results are compared with accurate multi-reference configuration interaction (MR-CI) calculations \cite{barbatti2004} and previous studies with neural-network VMC  \cite{entwistle2023, pfau2024}. 
Figure \ref{fig:ethylene} shows the energies of the ground state and the first excited state relative to the ground-state energy at the equilibrium geometry.
Our results are in excellent agreement with the NES-VMC method,  representing a significant improvement over previous calculations using penalty-based VMC \cite{entwistle2023}. 
The avoided crossing at $\tau=90^\circ$ is well reproduced, and the excitation energy of 2.41\,eV is within 0.01\,eV of the NES-VMC result.
Furthermore, we estimate the location of the conical intersection to be between 97.5$^\circ$ and 100$^\circ$ ($\sim 99^\circ$), bringing it closer to the MR-CI reference ($\sim 96^\circ$) and being in good agreement with the NES-VMC result($\sim 98^\circ$). 
We note that while the study of ethylene with NES-VMC required simulations for three excited states across the PES and a special treatment of the equilibrium geometry, we do not need to compute additional states and run all single-point calculations along the trajectory with the same parameters. 
The good agreement with NES-VMC and qualitative reproduction of the MR-CI results indicate that the penalty-based method is capable of accurately modeling the complicated electronic structure of the ethylene isomerization process.

\section{Discussion}
An updated version of the deep-learning penalty-based excited-state VMC method is presented and applied to compute a wide range of atomic and molecular excited states, demonstrating its enhanced accuracy and attractive computational properties.
The improvements include the use of a new state-of-the-art attention-based neural network ansatz, systematic tuning of the optimizer hyperparameters, and an updated overlap penalty term that guarantees the global minimum of the loss function yields the exact solution of the electronic structure problem.
The method's dependence on the choice of free parameters is greatly reduced by a formulation that automatically adapts these parameters to the physical system under consideration.
Lastly, a new penalty term is introduced which, in combination with a spin-assigned ansatz, enables the targeting of specific spin states, significantly improving the computational efficiency in many applications.

The computational aspects of the penalty-based method are examined in relation to other prevalent VMC-based algorithms for computing molecular excited states.
Compared to NES-VMC, penalty-based methods exhibit favorable scaling with the number of computed electronic states, due to the former approach's need to model and sample an extended system of fermions.
On the other hand, a remaining weak dependence on the choice of scaling parameters and accumulating noise from the penalty terms in rare, highly degenerate systems can cause comparatively worse numerical issues for the penalty-based method in a limited number of cases.
In contrast to variance minimization approaches, penalty-based methods do not require additional approximations to obtain stable gradients.
The approach presented here is more practical than both the NES-VMC algorithm, as it requires no additional diagonalization to recover the individual states, and the sequential variant of the penalty-based method, as it replaces a set of sequential calculations with a single parallel one, granting access to all electronic states at every stage of the computation.

The accuracy of results obtained with the penalty-based method is compared to both accurate reference values and numbers obtained with other deep-learning VMC-based approaches.
On a set of first- and third-row atoms, where practically exact experimental references are available, penalty-based excited-state VMC recovers the lowest nine excitation energies with a mean absolute error of less than 1 kcal/mol.
Its accuracy is on par with its sequential variant, and is comparable to that of the NES-VMC method, even though fewer training iterations are carried out here.
For fifteen small to medium molecules, deviations from theoretical reference energies remain well under control, while the reliably accurate oscillator strengths indicate consistently high wave function quality across a wide range of systems.
Considering excited-state potential energy surfaces, the efficiency and black-box nature of the penalty-based method in tackling this traditionally challenging class of problems become evident.
Large sections of nine carbon dimer potential energy surfaces with singlet, triplet, and quintet spins between bond lengths of 1.0 and 1.9 \AA{} are accurately recovered.
The derived vertical and adiabatic excitation energies are in good agreement with both experimental and theoretical reference values.
Turning to the ethylene potential energy surfaces, the updated penalty-based method significantly improves on results obtained with earlier versions and predicts the position of the conical intersection in good agreement with MR-CI and NES-VMC.
The description of both the carbon dimer and ethylene potential energy surfaces is made significantly more efficient by the use of the spin penalty term, roughly halving the number of states needed to be considered simultaneously in any single computation.

Overall, the computational advantages and accurate predictions of deep-learning penalty-based VMC place it among the most promising methods for computing electronic excited states with VMC.
Notably, it delivers similar accuracy to the NES-VMC method when the same neural-network ansatzes are used in both algorithms, while offering several practical advantages.
Given the recent surge in interest in deep-learning excited-state VMC for molecular electronic structure, the penalty-based approach is poised to become a prominent method for describing the most challenging excited states of small to medium molecules.

\section*{Acknowledgments}
We thank Alice Cuzzocrea and Paolo Erdman for helpful discussions.
Funding is gratefully acknowledged from Deutsche Forschungsgemeinschaft (DFG, German Research Foundation) under Germany´s Excellence Strategy – The Berlin Mathematics Research Center MATH+ (EXC-2046/1, project ID: 390685689) projects (AA1-6, AA2-8) and Deutsche Forschungsgemeinschaft (DFG, German Research Foundation) project NO825/3-2.
The authors gratefully acknowledge the computing time granted by the Resource Allocation Board and provided on the supercomputer Lise at NHR@ZIB as part of the NHR infrastructure. The calculations for this research were conducted with computing resources under the project bec00266.

\printbibliography
\appendix
\section{Spin assigned wave functions in quantum Monte Carlo}\label{sec:VMC_wf_and_spin}
Electronic wave functions $\Phi(\mathbf{x}_0, ...,\mathbf{x}_N)$ are functions of spacial and spin variables $\mathbf{x}_i = (\mathbf{r}_i, \sigma_i)$.
Being fermions, electrons obey anti-symmetric statistics and electronic wave functions change sign upon the permutation of any two electrons:
\begin{equation}
    \Phi(...,\mathbf{x}_i,..., \mathbf{x}_j,...) = - \Phi(...,\mathbf{x}_j,...,\mathbf{x}_i,...) \, .
\end{equation}
If the studied Hamiltonian does not depend on spin explicitly, as is the case for the molecular Hamiltonian (\ref{eq:hamil}), it shares a common set of eigenstates with the total spin operator $\hat{S}^2$.
The Hamiltonian and the total spin operator (as well as one of its components) can then be diagonalized independently, giving rise to a set of spacial and spin eigenfunctions. 
The spin eigenfunctions correspond to eigenvalues $S$ (spin quantum number) and $m_S$ (magnetic quantum number) and are linear combinations of products of the single electron spin indicator functions $\alpha$ and $\beta$.
Because of the independence of the spacial and spin parts, the spin eigenfunctions have to be symmetric (anti-symmetric) upon exchange of particles, with the spacial wave function being anti-symmetric (symmetric), respectively.
Eigenfunctions with spin quantum number $S$ have a multiplicity of $2S+1$, that is there exist $2S+1$ degenerate solutions (singlet, doublet, triplet, ...), each with a different $m_S$ value.
For example, for a two particle system there are three symmetric (triplet) and one anti-symmetric (singlet) spin eigenfunctions:
\begin{align}\label{eq:spin_example}
\begin{split}
    &\zeta_{1,1}~~ = \alpha(1)\alpha(2) \\
    &\zeta_{1,0}~~ = \frac{1}{\sqrt{2}}( \alpha(1)\beta(2)+\beta(1)\alpha(2)) \\
    &\zeta_{1,-1} = \beta(1)\beta(2) \\
    &\zeta_{0,0}~~ = \frac{1}{\sqrt{2}}( \alpha(1)\beta(2)-\beta(1)\alpha(2))\,,
\end{split}
\end{align}
where the two subscripts stand for the spin and magnetic quantum numbers, respectively.

When addressing the spin-independent molecular Hamiltonian, within VMC, it's often advantageous to optimize the spin-assigned spacial wave function rather than the fully antisymmetric wave function.
This approach imposes an anti-symmetry constraint for like-spin electrons (the spin component is generally symmetric under the exchange of same-spin particles), while eliminating the necessity to enforce symmetry requirements on opposite-spin particles.
The convention is to assign spin-up to the first $N_\uparrow$ electrons and spin-down to the remaining $N-N_\uparrow = N_\downarrow$ electrons.
While this effectively gives rise to distinguishable particles, spin independent expectation values are not affected by the assignment and full symmetry can be restored by multiplying with the spin function and anti-symmetrizing.
It is worth noting that in the process of fixing the spin of the electrons the magnetic quantum number is also determined, as the difference between the number of spin up and spin down electrons. 
The spin quantum number on the other hand depends on the symmetry of the spacial and spin wave functions and is determined in the optimization process.
During variational optimization, the spin assigned wave function will hence converge to the ground state among the wave functions compatible with the fixed $m_S$ component.
For the two electron example, picking a wave function with an equal number of spin-up and spin-down electrons allows expressing wave functions with all possible values of $S$ ($\zeta_{1,0}$ and $\zeta_{0,0}$), while a wave function with two same-spin electrons restricts to the $S=1$ spin sector (either $\zeta_{1,1}$ or $\zeta_{1,-1}$).

\end{document}